\documentclass[apjl]{emulateapj}
\usepackage{apjfonts}
\usepackage{times,color}

\slugcomment{Draft version December 30, 2006}

\shorttitle{
MAGNETIZED DIFFUSE CLOUD FORMATION
}
\shortauthors{Inoue et al.}

\begin{document}

\title{
THE ROLE OF AMBIPOLAR DIFFUSION IN THE FORMATION PROCESS OF MODERATELY MAGNETIZED DIFFUSE CLOUDS
}
\author{Tsuyoshi Inoue\altaffilmark{1}, Shu-ichiro Inutsuka\altaffilmark{1}, and Hiroshi Koyama\altaffilmark{2}}
\altaffiltext{1}{Department of Physics, Graduate Scool of Science, Kyoto University, Sakyo-ku, Kyoto 606-8588, Japan; tsuyoshi@tap.scphys.kyoto-u.ac.jp}
\altaffiltext{2}{Department of Astronomy, University of Maryland}

\begin{abstract}
We investigate the dynamical condensation process in a magnetized thermally bistable medium.
We perform one-dimensional two fluid numerical simulations that describe the neutral and ionized components in the interstellar medium with purely transverse magnetic fields.
We find that the clouds that are formed as a consequence of the thermal instability always have a magnetic field strength on the order of a few $\mu$G irrespective of the initial strength.
This shows good agreement with the measurements of the magnetic field strength in diffuse clouds.
We show analytically that the final magnetic field strength in the clouds is determined by the balance between ambipolar diffusion and the accumulation of the magnetic field due to condensation.
\end{abstract}


\section{Introduction}
It is widely known that the low- and mid-temperature parts of the interstellar medium (ISM) are described as a thermally bistable medium that results from the balance of radiative cooling and heating (Field et al.1969; Wolfire et al. 1995, 2003).
The two stable phases are called the cold neutral medium (CNM), and the warm neutral medium (WNM).

Thermal instability (TI) is the most promising formation mechanism of the CNM (Field 1965).
Recently, many authors studied the dynamical condensation process of the ISM driven by TI.
Koyama \& Inutsuka (2000, 2002), \cite{IK}, and \cite{IKI} studied the turbulent CNM formation as a result of TI in the layer compressed by shock propagation.
Hennebelle \& P\'erault (1999, 2000), \cite{AH}, Heitsch et al. (2005, 2006), and Vazquez-Semadeni et al(2006a, b) studied an analogous process in supersonic conversing flows.
However, there are not as many studies that take into account the effect of the magnetic field on TI (Elmegreen 1997; Hennebelle \& P\'erault 2000; Lim et al. 2005; Hennebelle \& Passot 2006).

Measurements of the magnetic field in the CNM (diffuse clouds) from the Zeeman effect give values around 5 $\mu$G (Myers et al. 1995; Heiles \& Troland 2005).
It shows that thermal and magnetic pressures in the diffuse clouds are comparable (plasma beta: $\beta=p_{\rm thermal}/p_{\rm mag} \sim 1$).
Thus, the effect of magnetic field on TI is obviously important.

In the case of a one-dimensional condensation process with a magnetic field that is frozen in the fluid (ideal-MHD), magnetic pressure increases in proportion to the square of the gas density, if the direction of the fluid motion is not perfectly aligned with the magnetic field.
Thus, during condensation, magnetic pressure seriously affects TI.
\cite{HP00} show that magnetic pressure prevent TI, if the initial angle between the magnetic field and the fluid velocity is larger than $20^\circ-30^\circ$ with a few $\mu$G initial field strength.
Furthermore, even if the CNM is formed as a result of TI, the resulting CNM would be occupied by material with low $\beta$ (see Lim et al 2005), because perfect alignment between the magnetic field and the velocity field is rare.
This seems to contradict the observational suggestion that the magnetic field strength is independent of density in the range $0.1$ to $100$ cm$^{-3}$; i.e., the WNM and the CNM have roughly the same field strengths (Troland \& Heiles 1986).  

However, the aforementioned studies on TI that take into account the effect of the magnetic field do not include the realistic diffusion process of the magnetic field. 
In this Letter, to deal with the realistic process of ambipolar diffusion during TI, we perform one-dimensional two-fluid numerical simulations with a purely transverse magnetic field.

\section{The Model}
\subsection{Basic Equations and Numerical Scheme}
We consider the equations of one-dimensional hydrodynamics and magnetohydrodynamics that have purely transverse magnetic fields, including collisional coupling between neutral and ionized gases (Draine 1986).
The neutral fluid part also includes radiative cooling, heating, and thermal conduction. 
\begin{eqnarray}
&& \frac{\partial\rho_{n,i}}{\partial t}+\frac{\partial}{\partial x}(\rho_{n,i}\,v_{n,i})=S_{n,i} ,\\
&& \frac{\partial}{\partial t}(\rho_{n,i}\,v_{n,i})+\frac{\partial}{\partial x}(\rho_{n,i}\,v_{n,i}^{2}+P_{n,i})=F_{n,i} ,\\
&& \frac{\partial E_{n,i}}{\partial t}+\frac{\partial}{\partial x}(E_{n,i}+P_{n,i})\,v_{n,i}=D_{n,i}+G_{n,i} ,\\
&& \frac{\partial B}{\partial t}=-\frac{\partial}{\partial x}(v_{i}\,B) ,\\
&& S_{n}=\alpha\,\frac{m_{\rm n}}{m_{\rm i}^{2}}\,\rho_{i}^{2}-\xi\,\rho_{n} , \,\,
 S_{i}=-\frac{\alpha}{m_{\rm i}}\,\rho_{i}^{2}+\frac{m_{\rm i}}{m_{\rm n}}\,\xi\,\rho_{n} , \\
&& F_{n}=-F_{i}=A\,\rho_{n}\,\rho_{i}\,(v_{i}-v_{n}) , \\
&& G_{n}=\frac{A\,\rho_{n}\,\rho_{i}}{m_{n}+m_{i}}\left\{ m_{i}(v_{i}-v_{n})^{2}+3\,k_{\rm B}(T_{i}-T_{n}) \right\} , \\
&& G_{i}=\frac{A\,\rho_{n}\,\rho_{i}}{m_{n}+m_{i}}\left\{ m_{n}(v_{i}-v_{n})^{2}+3\,k_{\rm B}(T_{n}-T_{i}) \right\} , \\
&& D_{n}=\frac{\partial}{\partial x}\left(\kappa\,\frac{\partial T_{n}}{\partial x}\right)-\rho_{n}\,\mathcal{L}(\rho_{n},T_{n}),\,\,D_{i}=0,
\end{eqnarray}
where the subscripts $n$ and $i$ denote the neutral and ion components, respectively.
In addition, we impose the ideal gas equation of state $p_{n,i}=k_{\rm B}\rho_{n,i}T_{n,i}/m_{n,i}$.
$P_{n}=p_{n}$ and $P_{i}=p_{i}+B^{2}/8\pi$ represent the thermal and magnetic pressure, and $E_{n}=p_{n}/(\gamma-1)+\rho_{n}v_{n}^{2}/2$ and $E_{i}=p_{i}/(\gamma-1)+\rho_{i}v_{i}^{2}/2+B^{2}/8\pi$  represent the total energy. 
We use the specific heats ratio $\gamma=5/3$ and the mean particle masses $m_{n}=m_{i}=1.27\,m_{\rm p}$.
The drag coefficient $A$, the recombination coefficient $\alpha$, the ionization coefficient $\xi$, and the thermal conductivity $\kappa$ are chosen to simulate the typical ISM whose properties are listed in Table 1.

We use the operator-splitting technique for solving the equations: the second-order Godunov method (van Leer 1979) for the inviscid fluid parts and the second-order explicit time integration for the cooling, heating and thermal conduction parts.
The parts of ionization, recombination, and frictional forces/heatings are updated using piecewise exact solutions for each time interval.

\begin{deluxetable}{clc}
\tablewidth{0pt}
\tablecaption{Sources}
\tablehead{
coefficient &  note  & reference
}
\startdata
$A(T)$ & $3.4 \times 10^{15}\left( T/8000\mbox{ K} \right)^{0.375}$ cm$^{3}$g$^{-1}$s$^{-1}$ & 1 \\
 & H$^{+}$+H collisions &  \\
$\kappa(T)$ & $2.5 \times 10^{3} \, T^{0.5}$ erg cm$^{-1}$s$^{-1}$K$^{-1}$ & 2 \\
 & H+H collisions & \\
$\alpha(T)$ & H$^{+}$ and He$^{+}$ recombinations & 3 \\
$\xi$ & H and He ionizations & 4 \\
 & due to CRs and X-ray &  \\
\enddata
\tablerefs{
(1) Glassgold et al. 2005; (2) Parker 1953; (3) Shapiro \& Kang 1995; (4) Wolfire et al. 1995.
}
\end{deluxetable}

\subsection{Cooling Function and Numerical Settings}
We take the following simplified net cooling function which approximates the realistic cooling function:
\begin{eqnarray}
\rho\,\mathcal{L} &=& n\,(-\Gamma +n\,\Lambda)\,\mbox{ erg cm}^{-3}\mbox{ s}^{-1},\\
\Gamma &=& 2\times 10^{-26},\\
\frac{\Lambda}{\Gamma} &=& 1.0\times 10^{7}\,\exp\left( \frac{-118400}{T+1000} \right) +
 1.4\times 10^{-2}\,\sqrt{T}\,\exp\left( \frac{-92}{T} \right).
\end{eqnarray}
This is obtained by fitting to the various heating ($\Gamma$) and cooling ($\Lambda$) processes considered by \cite{KI00}.
The resulting thermal-equilibrium state defined by $\mathcal{L}(n,T)=0$ is shown in the bottom panel of Figure \ref{f1} as a thick solid line. 

In all of the computations shown below, we use $N_{\rm cell}=4096$ grid points covering a $L=4$ pc region.
Thus, the spatial resolution is $\Delta x=L/N_{\rm cell} \simeq 10^{-3}$ pc, which satisfies the Field condition (Koyama \& Inutsuka 2004).
We impose a periodic boundary condition.
The initial condition of the neutral gas is static and uniform thermally unstable equilibrium whose gas pressure is $p_{n}/k_{\rm B}=3500$ K cm$^{-3}$, and that of the ionized gas is determined from the ionization equilibrium (in this case, $X_{\rm i}=n_{\rm i}/n_{\rm n}=3.10\times 10^{-2}$) with the same temperature as the neutrals.
In order to excite TI, we add density fluctuations to the neutral gas as
\begin{equation}
\delta \rho_{\rm n} = \mathcal{A}\,\sum_{l=1}^{l_{\rm max}}\,\,\sin\left[\,\frac{2\pi\,l}{L}x+\theta(l)\,\right],
\end{equation}
where $\theta(l)$ is a random phase.
We use $\mathcal{A}=0.05\, n_{\rm n,ini}$ and $l_{\rm max}=20$ in all runs.

As shown in \cite{HP00}, the transverse component of the magnetic field strongly affects the non-linear evolution of TI, even if the initial angle between the magnetic and velocity fields is small.
In order to study the nonlinear effect of magnetic pressure and ambipolar diffusion, we initially set a purely uniform transverse magnetic field whose strength is $B_{\rm ini}=0,\,0.03,\,0.1,\,0.3,$ and $1.0$ $\mu$G.
To contrast the effect of ambipolar diffusion, we also perform the ideal MHD simulation, which has the same cooling function and thermal conduction with $B_{\rm ini}=1.0$ $\mu$G.

\begin{figure}
\epsscale{1.0}
\plotone{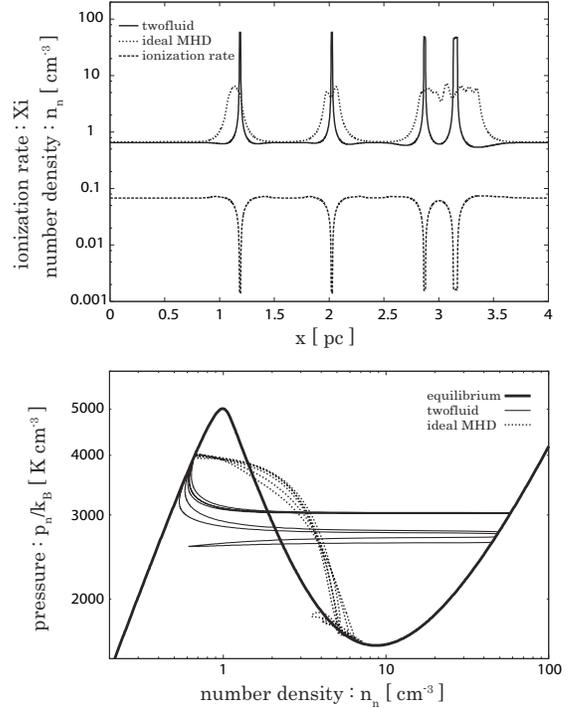}
\caption{
Density structure (top) and number density-pressure distribution (bottom) of the two-fluid case (thin solid line) and the ideal MHD case (dotted line) with $B_{\rm ini}=1.0$ $\mu$G at $t=25.0$ Myr.
The dashed line in the top panel shows the resulting ionization rate $X_{\rm i}=n_{\rm i}/n_{\rm n}$ of the two-fluid calculation.
}
\label{f1}
\end{figure}

\section{Results}
\subsection{The density evolution}
In all runs except the ideal MHD case, TI can fully develop and can create a two-phase medium (CNM and WNM).
On the other hand, in the ideal MHD case, the collapse of the dense region stops before the density reaches that of the thermally stable CNM branch because the magnetic pressure becomes strong enough to support the cold region.
Figure \ref{f1} shows the resulting density structure (top panel) and number density-pressure distribution (bottom panel) of the two-fluid case (thin solid line) and the ideal MHD case (dotted line) with $B_{\rm ini}=1.0$ $\mu$G at $t=25.0$ Myr.

\begin{figure}
\epsscale{1.0}
\plotone{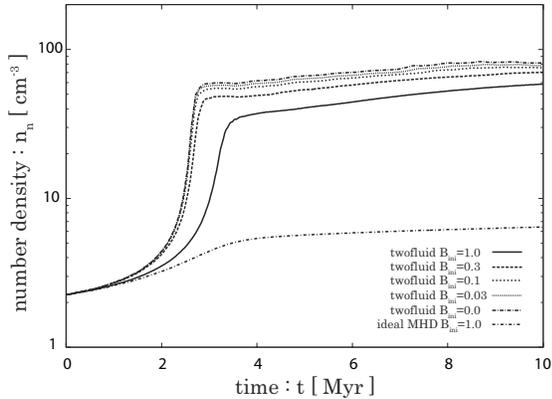}
\caption{
Neutral gas density evolutions of the densest point in the range $x\in [0.0,\,1.5]$ pc.
}
\label{f2}
\end{figure}

In the case of two-fluid calculation, the density evolution of the neutral gas is similar to that reported in \cite{KI06}.
At first, overdensity perturbations grow to the CNM phase.
The most unstable scale of the perturbation is written as $l_{\rm TI}\simeq\sqrt{l_{\rm a}\,l_{\rm F}}$ (Field 1965) which is $\sim 1$ pc in the unstable equilibrium and $0.1$ pc in the CNM, where $l_{a}\equiv c_{\rm s}\,t_{\rm cool}$ is an acoustic scale (the sound traveling scale within a cooling time) and $l_{\rm F}\equiv \sqrt{\kappa\,T/n^{2}\Lambda}$ is the Field length.
The growth timescale of the condensation is determined by the cooling time $t_{\rm grow}\sim t_{\rm cool}\equiv k_{\rm B}\,T/n\Lambda/(\gamma-1)\sim 1$ Myr in the unstable equilibrium and $\sim 0.1$ Myr in the CNM.
The top panel of Figure \ref{f2} shows the neutral gas density evolutions of the densest point in the range $x\in [0.0,\,1.5]$ pc, which contains only one evolving density peak for all of the runs.
Clearly, TI in the two-fluid case grows as fast as TI in the unmagnetized case ($B_{\rm ini}=0$).
This shows that ambipolar diffusion effectively causes the drifting of the magnetic field from the condensing region.

After the growth of the density enhancement, less dense regions grow into the WNM phase, because the growth timescale of the rarefaction is determined by the heating time, which is longer than the cooling time ($t_{\rm heat}\sim 1$ Myr in the unstable equilibrium and $\sim 10$ Myr in the WNM).
In the top panel of Figure 2 ($t=25$ Myr), we can see the WNM as the intercloud medium whose scale is $\sim 1$ pc, whose temperature is $\sim 8000$ K, and whose number density is $\sim 0.6$ cm$^{-3}$.

\subsection{The pressure evolution}
During the evolution, the total pressure $p_{\rm tot}=p_{\rm n}+p_{\rm i}+B^{2}/8\pi$ is spatially uniform, owing to the fact that the most unstable scale of TI is smaller than the acoustic scale $l_{\rm TI}<l_{\rm a}$.
The condensation terminates when the neutral gas reaches the stable phase or when the magnetic pressure becomes larger than the neutral gas pressure.
In our calculations, all two-fluid runs corresponds to the former case, and the ideal MHD run corresponds to the later case.

\subsection{The magnetic field strength evolution}
\begin{figure}
\epsscale{1.0}
\plotone{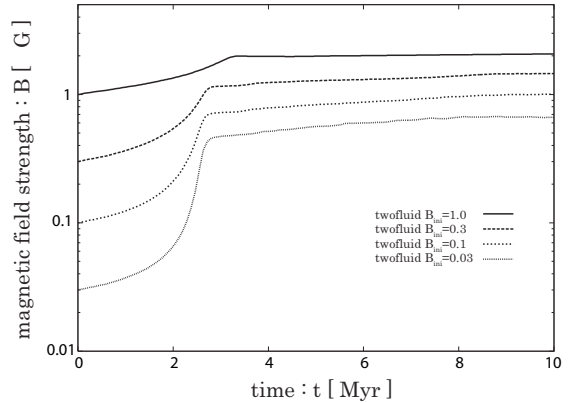}
\caption{
Magnetic field strength evolutions of the densest point in the range $x\in [0.0,\,1.5]$ pc.
}
\label{f3}
\end{figure}

Figure \ref{f3} shows the magnetic field strength evolutions of the densest point in the range $x\in [0.0,\,1.5]$ pc.
Amazingly, the magnetic field strength in the CNM always becomes an order of a few $\mu$G irrespective of the initial strength!
The resulting $\beta$'s in that point at $t=10$ Myr are $\beta = 2.46,\,5.55,\,12.2,$ and $28.2$, which correspond to the runs $B_{\rm ini}=1.0,\,0.3,\,0.1,$ and $0.03$ $\mu$G, respectively.

\section{Critical Magnetic Field Strength}
In this section, we derive the critical magnetic field strength at which ambipolar diffusion becomes effective.
We can estimate the condensation speed of TI as
\begin{equation} \label{vTI}
v_{\rm TI}\sim \frac{l_{\rm TI} }{t_{\rm cool}},
\end{equation}
where the spatial scale and the timescale are essentially determined by the most unstable scale of TI and the cooling time, respectively.
The drift speed of the magnetic field line from the condensing region can be estimated as follows.
During the condensation the ionization degree is small ($X_{\rm i}=10^{-2}-10^{-3}$); thus, the friction force and the Lorentz force are in balance.
In this case, the drift speed can be written as
\begin{equation} \label{vdrift}
v_{\rm drift}\equiv |\vec{v}_{\rm i}-\vec{v}_{\rm n}|=\frac{|(\vec{\nabla} \times \vec{B})\times \vec{B}|}{4\pi A\,\rho_{\rm n}\,\rho_{\rm i}}\sim \frac{B^{2}}{4\pi A\,X_{\rm i}\,\rho_{\rm n}^{2}\,l_{\rm TI}},
\end{equation}
(Nakano 1984; Shu 1992), where we simply replace the spatial derivative with the inverse of the most unstable scale of TI, which is the characteristic scale of the system.
If the drift speed is slower than the condensation speed $v_{\rm drift}<v_{\rm TI}$, the magnetic field lines are also condensed with the neutrals.
Such a case is realized when the magnetic field is weak.
If the drift speed is faster than the condensation speed $v_{\rm drift}>v_{\rm TI}$, the magnetic fieldlines drift from the condensing region.
Such a case is realized when the magnetic field is strong.
Therefore, during the evolution, the magnetic field strength inside the condensing region is adjusted to $v_{\rm drift}\sim v_{\rm TI}$.

Equating equations (\ref{vTI}) and (\ref{vdrift}), the critical strength at which the ambipolar diffusion becomes effective is written as follows:
\begin{equation}
B_{\rm AD}=\sqrt{4\pi\,A\,X_{\rm i}\,\rho_{\rm n}^{2}\,c_{\rm s}\,l_{\rm F}} \label{Bad1}.
\end{equation}
If we assume ionization equilibrium and isobaric evolution, equation (\ref{Bad1}) is written as a function of the neutral number density, which is plotted in Figure \ref{f3}.
Note that the ionization equilibrium is not maintained during the condensation, because the timescales of the recombination and ionization are, respectively, slightly smaller and larger than the cooling time.
Thus, this is an overestimation of the critical value during condensation.
This figure shows that the critical strength $B_{\rm AD}$ is not sensitive to density.
Therefore, we can express the value of the critical strength using typical values in the CNM:
\begin{equation}
B_{\rm AD}\simeq 3.2\Big( \frac{p_{\rm n}/k_{\rm B}}{4000\,\mbox{K cm}^{-3}} \Big)^{1.475}\Big( \frac{n_{\rm n}}{50 \,\mbox{cm}^{-3}} \Big)^{-0.4875}\mu\mbox{G} \label{Bad2} ,
\end{equation}
where we use $\xi\simeq 1.0\times 10^{-15}$ s$^{-1}$, $\alpha\simeq 6.8\times 10^{-11}\,T^{-0.5}$ cm$^{3}$ s$^{-1}$, and $\Lambda\simeq 3.6\times10^{-29}\,T^{0.8}$ erg cm$^{3}$ s$^{-1}$ (Wolfire et al. 2005).
This value shows good agreement with our numerical experiments and the measurements in the diffuse HI cloud, and this is why the resulting magnetic field strength in the CNM always converges to the order of a few $\mu$G irrespective of the initial strength.

\begin{figure}
\epsscale{1.0}
\plotone{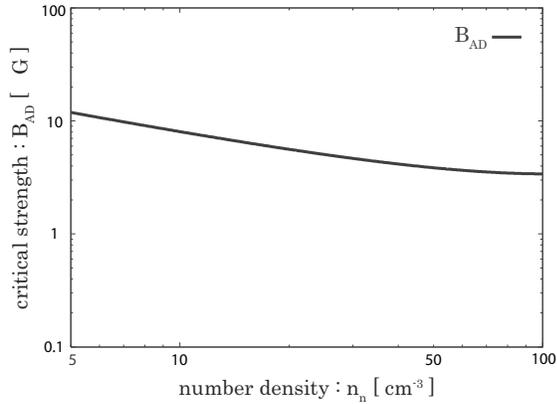}
\caption{
Critical magnetic field strength vs. number density of the neutrals.
}
\label{f4}
\end{figure}

\section{Summary and Discussions}
We studied the dynamical condensation process of TI, which takes into account the effect of magnetic pressure and ambipolar diffusion using a one-dimensional two-fluid numerical simulation.
We showed that even if $\beta \sim 1$ initially, TI can fully develop and generate a moderately magnetized cloud ($\beta \sim 1$) owing to ambipolar diffusion.
Therefore, if diffuse HI clouds are formed by TI, it is quite natural for the observations to show no increase in magnetic field strengths over the density range 0.1 to 100 cm$^{-3}$(Troland \& Heiles 1986).

Furthermore, we showed that the magnetic field strength in the CNM converges to the order of a few $\mu$G irrespective of the initial strength.
This result can be understood in the following: during the condensation, the magnetic field strength is tuned to the critical value owing to the accumulation of magnetic field lines by TI and ambipolar diffusion.
The critical value becomes around 3 $\mu$G in the typical ISM (see equation [\ref{Bad2}]).
This value shows good agreement with the measurements of the magnetic field strength in the diffuse HI clouds (Myers et al. 1995; Heiles \& Troland 2005).
Thus, our result suggest that even if there are large fluctuations in the magnetic field strength in the WNM, the strength in the diffuse clouds always becomes a few $\mu$G.

In this study, for simplisity, we chose a thermally unstable equilibrium to the initial condition.
A study of more realistic situation (for example, the converging of two flows or shock propagation) will be explored in our next paper. 
In the case of two flows converging, the ram pressure of the flows will prevent the ions from drifting.
Thus, ambipolar diffusion can become effective after the converging flows become weak.

\acknowledgments
This work is supported by the Grant-in-Aid for the 21st Century COE ``Center for Diversity and Universality in Physics" from the Ministry of Education, Culture, Sports, Science and Technology (MEXT) of Japan.

\end{document}